\documentstyle[times,pramana,epsf,floats,epsfig]{ias}

\newcommand\prl[3]   {{Phys.\ Rev.\ Lett.\ }{\bf #1} (#2) #3}
\newcommand\plb[3]   {{Phys.\ Lett.\ }{\bf B #1} (#2) #3}
\newcommand\npb[3]   {{Nucl.\ Phys.\ }{\bf B #1} (#2) #3}
\newcommand\prep[3]  {{Phys.\ Rept.\ }{\bf #1} (#2) #3}

\newcommand\prd[3]   {{Phys.\ Rev.\ }{\bf D #1} (#2) #3}
\newcommand\jhep[3]  {{JHEP\ }{\bf #1} (#2) #3}

\newcommand{\newc}{\newcommand}
\newc\eg{{\it {e.g.}}}  \newc\etal{{\it {et al.}}} \newc\ie{{\it i.e.}}
\newc\etc{{\it {etc}}}  \newc\ibid{{\it {ibid}}}
\newc\vs{{\it {vs.}}}

\newcommand\lsim{\mathrel{\rlap{\lower4pt\hbox{\hskip1pt$\sim$}}
    \raise1pt\hbox{$<$}}}
\newcommand\gsim{\mathrel{\rlap{\lower4pt\hbox{\hskip1pt$\sim$}}
    \raise1pt\hbox{$>$}}}

\newc{\sigsip}{\sigma^{SI}_{p}}	\newc{\sigsin}{\sigma^{SI}_{n}}
\newc{\sigsdp}{\sigma^{SD}_{p}}	\newc{\sigsdn}{\sigma^{SD}_{n}}

\newc\bsgamma{b\rightarrow s\gamma}
\newc\brbsgamma{BR(B\rightarrow X_s\gamma)}
\newc\brbsmumu{BR(B_s\rightarrow \mu^+ \mu^-)}

\newc{\mhalf}{m_{1/2}}      \newc{\mzero}{m_0}

\newc\sigmaew{\sigma_{EW}}
\newc\sigmaint{\sigma_{int}}

\newc\fa{f_a}
\newc\mw{m_W}		\newc\mz{m_Z}

\newc{\tanb}{\tan\beta}
\newc{\azero}{A_0}
\newc{\at}{A_t} \newc{\abot}{A_b} \newc{\atau}{A_\tau} 

\newc{\bmu}{B\mu}           \newc{\sgn}{{\rm sgn}}
\newc{\mone}{M_1}           \newc{\mtwo}{M_2}

\newc{\charone}{\chi_1^\pm} \newc{\mcharone}{m_{\chi_1^\pm}}

\newc{\hl}{h}               \newc{\mhl}{m_{\hl}}
\newc{\hh}{H}               \newc{\mhh}{m_{\hh}}
\newc{\ha}{A}               \newc{\mha}{m_{\ha}}
\newc{\hc}{H^{\pm}}         \newc{\mhc}{m_{\hc}}

\newc{\qzero}{Q_0}          \newc{\qstop}{Q_{\widetilde t}}

\newc{\amu}{a_{\mu}}        \newc{\amususy}{a_{\mu}^{\rm SUSY}}
\newc{\amuexpt}{a_{\mu}^{\rm expt}}        \newc{\amusm}{a_{\mu}^{\rm SM}}
\newc{\deltaamususy}{\Delta a_{\mu}^{\rm SUSY}}

\newc{\msbar}{\overline {\rm MS}} \newc{\drbar}{\overline {\rm DR}}

\newc{\mt}{m_t} \newc{\mb}{m_b} \newc{\mtau}{m_{\tau}}
\newc{\yt}{h_t} \newc{\yb}{h_b} \newc{\ytau}{h_{\tau}}
\newc{\mtpole}{m_t^{\rm pole}} \newc{\mbpole}{m_b^{\rm pole}} 
\newc{\mtaupole}{m_{\tau}^{\rm pole}} 

\newc{\mtmtsmmsbar}{m_t(m_t)^{\msbar}_{{\rm SM}}}
\newc{\mtmtsmdrbar}{m_t(m_t)^{\drbar}_{{\rm SM}}}
\newc{\mtmtmssmdrbar}{m_t(m_t)^{\drbar}_{{\rm SUSY}}}

\newc{\mbmbsmmsbar}{m_b(m_b)^{\msbar}_{{\rm SM}}}
\newc{\mbmzsmmsbar}{m_b(\mz)^{\msbar}_{{\rm SM}}}
\newc{\mbmzsmdrbar}{m_b(\mz)^{\drbar}_{{\rm SM}}}
\newc{\mbmzmssmdrbar}{m_b(\mz)^{\drbar}_{{\rm SUSY}}}
\newc{\mtaumzsmmsbar}{m_{\tau}(\mz)^{\msbar}_{{\rm SM}}}
\newc{\mtaumzsmdrbar}{m_{\tau}(\mz)^{\drbar}_{{\rm SM}}}
\newc{\mtaumzmssmdrbar}{m_{\tau}(\mz)^{\drbar}_{{\rm SUSY}}}

\newc{\mgut}{M_{\rm GUT}}

\newc{\mplanck}{M_{\rm P}}      \newc{\mpl}{M_{\rm Pl}}
\newc{\msusy}{M_{\rm SUSY}}      \newc{\ms}{M_{\rm S}}

\newc{\jxf}{J({\xf})}
\newc{\jxfexact}{J_{\rm exact}({\xf})}  \newc{\jxfexp}{J_{\rm exp}({\xf})}
\newc{\VEV}[1]{\langle #1 \rangle}

\newc{\xf}{x_f}
\newc\vrel{v_{\rm rel}}
\newcommand\mchi{m_{\chi}}              

\newc\sell{{\widetilde e}_L}      \newc\msell{m_{\sell}}
\newc\selr{{\widetilde e}_R}      \newc\mselr{m_{\selr}}

\newc\snu{{\widetilde \nu}}

\newc\snue{{\widetilde \nu}_e}      \newc\msnue{m_{\snue}}
\newc\snutau{{\widetilde \nu}_\tau}      \newc\msnutau{m_{\snutau}}

\newc\supl{{\widetilde u}_L}      \newc\msupl{m_{\supl}}
\newc\supr{{\widetilde u}_R}      \newc\msupr{m_{\supr}}
\newc\sdl{{\widetilde d}_L}      \newc\msdl{m_{\sdl}}
\newc\sdr{{\widetilde d}_R}      \newc\msdr{m_{\sdr}}

\newcommand\stopq{{\widetilde t}}   \newcommand\mstopq{m_{\stopq}}

\newcommand\stopone{{\widetilde t}_1}   \newcommand\mstopone{m_{\stopone}}
\newcommand\stoptwo{{\widetilde t}_2}   \newcommand\mstoptwo{m_{\stoptwo}}

\newcommand\stauone{{\widetilde \tau}_1}

\newcommand\mgluino{m_{\widetilde g}}

\newc\hpm{H^\pm} \newc\hp{H^+} \newc\hm{H^-} 
\newc\sfermion{\tilde f}  \newc\msfermion{m_{\sfermion}}  

\newc\second{{\rm sec}} 

\newc\alphas{\alpha_s}

\newc\alphaem{\alpha_{em}}

\newc{\gstar}{g_\ast}           \newc{\gsstar}{g_{s\ast}}
\newc{\geff}{g_{\rm eff}}

\newcommand\treh{T_{\rm R}}

\newc{\sthw}{\sin\theta_W}              \newc{\cthw}{\cos\theta_W}
\newc{\bino}{\widetilde B}              \newc{\wino}{\widetilde W_3}
\newc{\higgsinob}{{\widetilde H}^0_b}   \newc{\higgsinot}{{\widetilde H}^0_t}

\newc{\abund}{\Omega h^2}
\newc{\abundchi}{\Omega_\chi h^2}
\newc{\abundcdm}{\Omega_{{\rm CDM}} h^2}
\newc{\omegam}{\Omega_{{\rm M}}}       \newc{\abundm}{\Omega_{{\rm M}} h^2}
\newc{\omegab}{\Omega_{{\rm b}}}	\newc{\abundb}{\Omega_{{\rm b}} h^2}
\newc{\omegacdm}{\Omega_{{\rm CDM}}}   \newc{\omegatot}{\Omega_{{\rm TOT}}}

\newc{\rhocrit}{\rho_{crit}}
\newc{\rhochi}{\rho_{\chi}}

\newcommand\tev{\,\mbox{TeV}}
\newcommand\gev{\,\mbox{GeV}}

\newcommand\kev{\,\mbox{keV}}
\newcommand\ev{\,\mbox{eV}}

\newcommand\cm{\,\mbox{cm}}
\newcommand\pb{\,\mbox{pb}}

\newc\br{\mbox{BR}}

\newc{\ra}{\rightarrow}
\newc{\beq}{\begin{equation}}
\newc{\eeq}{\end{equation}}
\newc{\bea}{\begin{eqnarray}}
\newc{\eea}{\end{eqnarray}}

\renewcommand\({\left(}
\renewcommand\){\right)}

\begin{document}
\mark{{L.~Roszkowski, Particle Dark Matter}{L.~Roszkowski, Particle Dark Matter}}
\title{Particle Dark Matter - A Theorist's Perspective\footnote{
Invited talk at PASCOS--03, Mumbai, India, 3--8 January 2003.}
}

\author{Leszek Roszkowski} 
\address{Department of Physics and
Astronomy, University of Sheffield, Hicks Building, Sheffield, S3 7RH,
England} 
\pacs{2.0}
\abstract{Dark matter is presumably made of some new, exotic particle
that appears in extensions of the Standard Model. After giving a brief
overview of some popular candidates, I discuss in more detail the most
appealing case of the supersymmetric neutralino.  }

\maketitle
\section{Introduction -- WIMP--Type Candidates for DM}
While one cannot complain about a shortage of candidates for
explaining the nature of the dark matter (DM) in the Universe, from
the point of view of particle physics, the WIMP (weakly interacting
massive particle) looks particularly attractive. In many ``scenarios''
as well as more complete theories beyond the SM there often appear
several new WIMPs and it is typically not too difficult to ensure that
the lightest of them is stable by means of some discrete symmetry or
topological  invariant. (For example, in supersymmetry, one invokes
$R$--parity.) In order to meet stringent astrophysical
constraints on exotic relics (like anomalous nuclei), they must be
electrically and (preferably) color neutral. They can however interact
weakly.

Contrary to common wisdom, WIMP--type candidates are neither bound to
interact with roughly weak interaction strength (in the sense of
electroweak) nor does their mass have to be in the $\gev$ to $\tev$
regime. From a particle theorist's point of view, it is most sensible to
concentrate on the cases where the WIMP appears as a ``by--product''
in some reasonable frameworks beyond the SM which have been invented to
address some other major puzzle in particle physics. In other words, let's
talk about WIMP candidates that have not been invented for the sole
purpose of solving the DM problem.

\begin{figure}[t!]
\epsfxsize=8cm
\centerline{\epsfbox{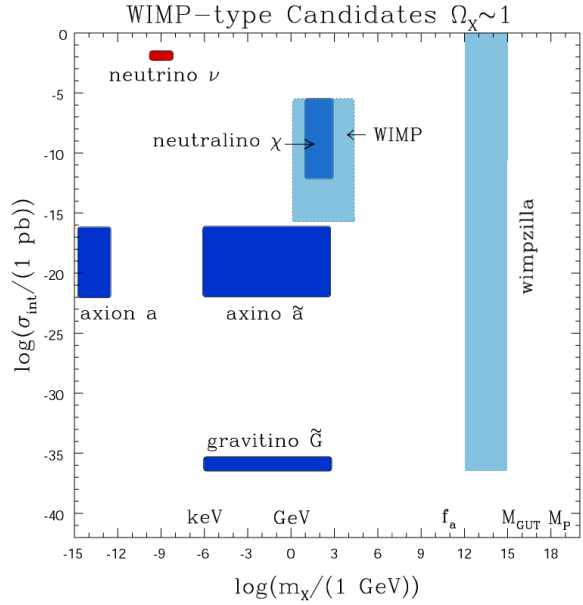}}
\caption{A schematic representation of some
well--motivated WIMP--type particles for which a priori one can have
$\omega\sim1$. $\sigma_{\rm int}$ represents a typical order of
magnitude of interaction strength with ordinary matter. The neutrino
provides hot DM which is disfavored. The box marked ``WIMP' stands for
several possible candidates, \eg, from Kaluza--Klein scenarios.
}
\label{bigpic:fig}
\end{figure}

One way to present this is to consider a big ``drawing board'' as in
Fig.~\ref{bigpic:fig}: a plane spanned by the mass of the WIMP on the
one side and by a typical strength $\sigmaint$ of its interaction with
ordinary matter (\ie, detectors) on the other. To a first
approximation the mass range can in principle extend up to the Planck
mass scale, but not above, if we are talking about elementary
particles. The interaction cross section could reasonably be expected
to be of the electroweak strength 
($\sigmaew \sim 10^{-38}\cm^2=10^{-2}\pb$) but could also be as tiny
as that purely due to gravity:
$\sim\(\mw/\mplanck\)^2\sigmaew\sim10^{-32}\sigmaew\sim10^{-34}\pb$.

What can we put into this vast plane shown in Fig.~\ref{bigpic:fig}?
One obvious candidate is the neutrino, since we know that it exists.
Neutrino oscillation experiments have basically convinced us that its
mass of at least $\sim0.1\ev$. On the upper side, if it were heavier
than a few $\ev$, it would overclose the Universe. The problem of
course is that such a WIMP would constitute {\em hot} DM which is
hardly anybody's favored these days. While some like it hot, or warm,
most like it cold.

Cold, or non--relativistic at the epoch of matter dominance (although
not necessarily at freezeout!) and later, DM particles are strongly
favored by a few independent arguments. One is numerical simulations
of large structures. Also, increasingly accurate studies of CMB
anisotropies, most notably recent results from WMAP~\cite{wmap0302},
imply a large cold DM (CDM) component and strongly suggest that most
($\sim90\%$) of it is non--baryonic.

In the SUSY world, of course we could add a sneutrino $\snu$, which,
like neutrinos, interacts weakly. From LEP its mass $\gsim70\gev$
(definitely a cold DM candidate), but then $\Omega_{\snu}\ll
1$. Uninteresting and $\snu$ does not appear in Fig.~\ref{bigpic:fig}.

The main suspect for today is of course the neutralino
$\chi$. Unfortunately, we still know little about its properties.  LEP
bounds on its mass are actually not too strong, nor are they robust:
they depend on a number of assumptions.  In minimal SUSY (the
so-called MSSM) `in most cases' $\mchi\gsim70\gev$, but the bound can
be also much lower.  Theoretically, because of the fine tuning
argument, one expects its mass to lie in the range of several tens or
hundreds of $\gev$. More generally,  $\mchi\gsim
{\rm few}\gev$ from $\abundchi\lsim1$ (the so--called Lee-Weinberg
bound~\cite{leeweinberg77}) and $\mchi\lsim300\tev$ from
unitarity~\cite{gk90}. Neutralino interaction rates are generally
suppressed relative to $\sigmaew$ by various mixing angles in the
neutralino couplings. In the MSSM they are typically between
$\sim10^{-3}\sigmaew$ and $\sim10^{-10}\sigmaew$, although could be
even lower in more complicated models where the LSP would be dominated
for example by a singlino (fermionic partner of an additional Higgs
singlet under the SM gauge group). This uncertainty of the precise
nature of the neutralino is reflected in Fig.~\ref{bigpic:fig} by
showing both the smaller (dark blue) region of minimal SUSY and an
extended one (light blue) with potentially suppressed interaction
strengths in non--minimal SUSY models. Another example of a WIMP that
would belong to the light blue box is the the lightest Kaluza--Klein
state which is massive, fairly weakly interacting and stable in some
currently popular Kaluza--Klein frameworks~\cite{st02}.

One can see that, while the (s)neutrinos interact  weakly {\em sensu stricto},
this is not quite the case with the neutralino and the other WIMP
candidates above. In fact, a typical strength of their interactions can
be several orders of magnitude less, while still giving $\abund\sim 1$. 

There are some other cosmologically relevant relics out there whose
interactions would be much weaker than electroweak.  One well--known
know example is the axion -- a light neutral pseudoscalar particle
which is a by--product of the Peccei--Quinn solution to the strong CP
problem. Its interaction with ordinary matter is suppressed by the PQ
scale $\sim(\mw/\fa)^2\sigmaew\sim10^{-18}\sigmaew\sim10^{-20}\pb$
($\fa\sim10^{11}\gev$), hence extremely tiny, while its mass $m_a\sim
\Lambda_{QCD}^2/\fa\sim \left( 10^{-6}-10^{-4}\right)\ev$ if
$\Omega_a\sim1$. The axion, despite being so light, is of CDM--type
because it is produced by the non--thermal process of misalignment in
the early Universe.

In the supersymmetric world, the axion has its fermionic superpartner,
called axino. Its mass is strongly model--dependent but, in contrast
to the neutralino, often not directly determined by the SUSY breaking
scale $\sim1\tev$. Hence the axino could be light and could naturally
be the LSP, thus stable. An earlier study concluded that axinos could
be {\em warm} DM with mass less than $2\kev$~\cite{rtw90}. More
recently it has been pointed out more massive axinos quite naturally
can be also {\em cold} DM as well, as marked in
Fig.~\ref{bigpic:fig}. Relic axinos can be produced either through
thermal scatterings and decays involving gluinos and/or squarks in the
plasma, or in out-of-equilibrium decays of the next--to-LSP, \eg\ the
neutralino~\cite{ckrplus}. The first mechanism is more efficient at
larger reheat temperatures $\treh\gsim10^4\gev$, the second at lower
ones. Axino cosmology is very interesting but I have no time to
discuss it here~\cite{ckrplus}.

Lastly, there is the gravitino -- the fermionic superpartner of the
graviton -- which arises by coupling SUSY to gravity. The gravitino
relic abundance can be of order one~\cite{gravitinoproduction}  
but one has to also worry 
about the so--called gravitino problem: heavier particles, like the
NLSP, will decay to gravitinos very late, around $10^7\sec$ after the
Big Bang, and the associated energetic photons may cause havoc to BBN
products. The problem is not unsurmountable but more
conditions/assumptions need to be satisfied. 
One way is to assume that the NLSP is mostly a higgsino but this does
not normally happen in the framework of grand unification, at least
not in minimal models. 
Fig.~\ref{bigpic:fig} the gravitino is marked in the mass range of
$\kev$ to $\gev$ and gravitational interactions only, although
$\kev$--gravitinos have actually strongly enhanced couplings via their
goldstino component.

While Fig.~\ref{bigpic:fig} is really about WIMPs which arise in
attractive extensions of the SM, it is worth mentioning another class
of relics, popularized under the name of WIMPzillas, for which there
exist robust production mechanisms (curvature perturbations) in
the early Universe~\cite{wimpzilla:ref}. As the name suggests, they are thought to be very
massive, $\sim10^{13}\gev$ or so. There are no restrictions on
WIMPzilla interactions with ordinary matter, other than they must
interact at least gravitationally, as schematically depicted in
Fig.~\ref{bigpic:fig}.

In summary, the number of {\em well--motivated} WIMP and WIMP--type
candidates for CDM is in the end not so large. On the other hand, one
should remember that in the box marked with the generic name ``WIMP''
one can accommodate not just the neutralino but also some other stable
states appearing in various extensions of the SM, \eg, Kaluza--Klein
type theories. One can add to this picture other candidates, like
cryptons and other particles arising in the context of superstrings,
or mirror DM~\cite{mohapatra:mirrordm}.

SUSY, which to many is the most promising extension of the
SM, provides three robust WIMP candidates for the
CDM: the neutralino, the axino and the gravitino. Each has its virtues
and weak points but I will have no time to discuss this here.  I think
that it is fair to say that the WIMP for today, and this decade, is
the neutralino. It is therefore of our primary interest here because
it is present in any sensible SUSY spectrum and is testable in today's
experimental programmes.

In this talk, I will explore cosmological properties of the neutralino
as the LSP in the general MSSM and in two unification--based models:
the Constrained MSSM (CMSSM) and in the GUT model based on the $SO(10)$ gauge
group. (See the talk by Raby~\cite{raby_pascostalk}). I will next present ensuing predictions
for the cross section for DM neutralino elastic scattering with
detector material in both models and contrast them with the case of
the general MSSM. See also the complementary presentation of Nojiri~\cite{nojiri_pascostalk} for a
discussion of indirect detection and collider search for SUSY
aspects of the neutralino.

\section{The Neutralino}

\subsection{The Frameworks}

There are two basic schemes in which one usually considers
cosmological properties of the neutralino. One is a rather general
framework of the MSSM where superpartner masses originate primarily
from soft SUSY--breaking terms. Additionally, one assumes a common
mass parameter $\mhalf$ for all the gauginos in the spectrum which
leads to the well--known relations
$\mone=\frac{5}{3}\tan^2_{\theta{W}}\mtwo\simeq 0.5\mtwo$ and
$\mtwo=\frac{\alpha_2}{\alphas}\mgluino\simeq0.3\,\mgluino$, with
$\mone/\mtwo/\mgluino$ being the soft bino/wino/gluino mass. One
further imposes the $R$--parity to make the LSP stable.

Alternatively, it is more popular today to consider specific boundary
conditions at some high scale, like the grand unification scale
$\mgut$ or the string scale.  In unified models one
writes down the Lagrangian at the unification scale
and next employs the Renormalization Group Equations (RGEs) to compute
the couplings and masses in an effective theory valid at the
electroweak scale.  One such popular model is the CMSSM, aka mSUGRA,
where one assumes a common soft mass scale $\mzero$ for all the
scalars (sfermions and Higgs) and a common trilinear soft
SUSY--breaking parameter $\azero$. These parameters are run using
their respective RGEs down from $\mgut$ to some appropriately chosen
low-energy scale $\qzero$ where the Higgs potential (including full
one-loop corrections) is minimized while keeping the usual ratio
$\tanb=\frac{v_t}{v_b}$ of the Higgs VEVs fixed. The Higgs/higgsino
mass parameter $\mu$ and the bilinear soft mass term $\bmu$ are then
computed from the conditions of radiative electroweak symmetry
breaking (EWSB), and so are the Higgs and superpartner masses. The
CMSSM thus a priori has only the usual $\tanb,\ \ \mhalf,\ \ \mzero,\
\ \azero,\ \ \sgn(\mu)$ as input parameters.  However, in the case of
large $\mhalf,\mzero\gsim1\tev$ and/or large $\tanb\sim {\cal
O}(\mt/\mb) $ some resulting masses will in general be highly
sensitive to the assumed physical masses of the top and the bottom (as
well as the tau).

Another interesting scenario is a fully realistic effective SUSY model
which derives from a minimal GUT with the $SO(10)$ gauge group
(MSO$_{10}$SM~\cite{bdr,raby_pascostalk}). It involves a different set
of boundary condtions at $\mgut$ which leads to a distinctively
different set of phenomenological and cosmological
predictions~\cite{drrr1,raby_pascostalk}. On starts with a well
defined model at the GUT scale: the sfermions of all the three
families have a unified (soft) mass $m_{16}$, while for the Higgs
(which come in ${\bf 10_H}$ and its conjugate) the analogous unified
quantity is $m_{10}$. After running the parameters down to $\mz$, one
finds that experimental constraints require $A_0 \sim - 2 \; m_{16}$,
$m_{10} \sim \sqrt{2} \; m_{16}$, $m_{16} \geq 2\tev \gg\mu, \mhalf$,
and $\Delta m_H^2 \sim 10 \%$, where $m_{(H_t,\; H_b)}^2 = m_{10}^2 (
1 \mp \Delta m_H^2)$~\cite{bdr}.  $\tanb$ is necessarily large
$\sim50$ because of $t-b-\tau$ Yukawa unification.

The case of large $\tanb$ requires one to treat the top, bottom and
tau masses with special care. These (especially $m_b$) receive large
radiative corrections from SUSY and one has to be careful in
extracting from them the corresponding Yukawa couplings which in turn
have an important effect on the running of the RGEs at large $\tanb$.
The masses of the top and tau are treated with a similar accuracy
although corrections to their masses are typically smaller.

Despite a small number of independent parameters in unified models,
their analysis is often technically rather involved, especially at
large $\tanb$, as mentioned above. Furthermore, in order to reduce the
scale dependence of the Higgs sector and related conditions for the
EWSB it is important to include full one-loop corrections to the Higgs
potential {\em and} also minimize the Higgs potential at the scale
$\qstop\sim\sqrt{\mstopone \mstoptwo}$ with $\mstopone$ ($\mstoptwo$)
denoting the physical masses of the stops.  This is because, at this
scale, the role of the otherwise large log-terms
$\sim\log\left(\mstopq^2/Q^2\right)$ from the dominant stop-loops will
be reduced.  At this scale one evaluates the one-loop conditions for
the EWSB which determine $\mu^2$ as well as the bilinear soft mass
parameter $\bmu$. Alternatively, one can run down all the mass
parameters down to $\mz$ (or their physical values, if they are above
$\mz$) and include all the threshold corrections to the
$\beta$--functions due to the decoupling of states.

The mass of the pseudoscalar $\mha$ plays an important role in
evaluating $\abundchi$, especially at very large $\tanb\sim50$.  This
is so for three reasons: (i)~$\mha$ decreases with increasing
$\tanb$~\cite{dn93} due to the increased role of the bottom Yukawa
coupling which at large enough $\tanb$ opens up a wide resonance
$\chi\chi\ra f\bar{f}$; (ii)~because the $A$-resonance is dominant due to
the coupling $Af\bar{f}\sim \tanb$ for down-type fermions; and
(iii)~because, in contrast to the heavy scalar $\hh$, this channel is not
$p$-wave suppressed~\cite{jkg96:ref}.  One needs to stress however
that there still remains considerable uncertainty (of the order of
$\sim5-10\%$) in the procedure for computing Higgs masses and
conditions for EWSB.

\subsection{The Bino is the winner}

The lightest neutralino is the lowest--lying mass eigenstate of the
two gauginos ($\bino$ and $\wino$) and the two higgsinos
($\higgsinob,\higgsinot$) which are the fermionic superpartners of the
neutral gauge and Higgs boson states, respectively. Remarkably,
despite a priori many free parameters present in SUSY theories, the
state that appears to most naturally give $\abundchi\sim1$ is the
nearly pure bino~\cite{chiasdm}. This is true in both the MSSM and in
basically all unified models. Indeed, for the higgsino one typically
finds $\abundchi\ll1$ due to efficient annihilation to $WW,ZZ,{\bar
t}t$ and coannihilation. In unified models, the LSP neutralino is
often a nearly pure bino~\cite{na92,rr93,kkrw94:ref} because of the
requirement of radiative EWSB which typically gives $|\mu|\gg \mone$.
This often allows one to impose strong constraints from $\abundchi<
{\cal O}(1)$ on {\em most} of the region of $\mhalf$ and $\mzero$ in
the ballpark of $1\tev$, as originally shown in
Refs.~\cite{rr93,kkrw94:ref} and later confirmed by many
studies. There a few exceptions to this rough upper bound which play
an important role in unified SUSY models. The bino as the
cosmologically favored LSP has now become part of the standard lore.

\subsection{Experimental Constraints}\label{expt:sec}

First I'll
summarize the relevant experimental constraints.

{\bf 1.}~$\mcharone>104\gev$ from LEP. This constraint is fairly robust,
  except for some fairly degenerate cases.

{\bf 2.}~$\mhl>111\gev$ in the case of light SM--like Higgs which holds
  in unified models. The actual limit from LEP is $\mhl>114.1\gev$ but
  theoretical uncertainty is still about $2-3\gev$. In the MSSM, there
  remains a narrow corridor of $\mhl\simeq\mha$ down to some 90\gev.

{\bf 3.}~$\br( B\ra X_s \gamma)=(3.34\pm0.68)\times 10^{-4}$~\cite{rrn1,or1+2}. This
constraint is very important in the light of a rather good agreement
between experiment and 
the SM prediction (${\brbsgamma}_{SM} = ( 3.70 \pm 0.30 ) \times
10^{-4}$~\cite{bcmu02}). However, it is also 
very sensitive to underlying theoretical assumptions and
has to be applied with much care.  If one makes the usual simplifying
assumption that the mass mixing in the down--type squark sector is the
same as in the corresponding quark sector (the so--called minimal
flavor violation, or MFV, scenario), then the constraint provides a
very strong lower bound on $\mhalf$ and $\mzero$, especially at large
$\tanb$ where SUSY contributions as strongly enhanced. However, by
allowing for even a mild relaxation of the MFV assumption, one finds
that the constraint from $b\rightarrow s\gamma$ may become very much
weaker, or even disappear altogether~\cite{or1+2}. We will show its
effect in the (C)MSSM but not in the MSO$_{10}$SM where the MFV
scenario does not necessarily hold. The constraint favors $\mu>0$ although the are
ways to overcome this~\cite{or1+2}.

{\bf 4.}~$\amuexpt-\amusm=(22.1\pm11.3)\times10^{-10}$ ($e^+e^-$ data) and
$\amuexpt-\amusm=(7.4\pm10.5)\times10^{-10}$ ($\tau$ data), where
$\amu=(g_\mu-2)/2$ is the anomalous magnetic moment of the muon. These
numbers are quoted from the latest analysis~\cite{dehz03}. This
reduces previous discrepancy with the SM value: $1.9\,\sigma$ and
$0.7\,\sigma$, respectively, and also between the analyses using the
$e^+e^-$ and $\tau$--data. It also favors $\mu>0$.

{\bf 5.}~$0.095< \abundchi < 0.13$ ($2\,\sigma$), a stringent range determined
  recently by WMAP (+CBI and ACBAR)~\cite{wmap0302}. Larger values are
  excluded. Smaller ones are in principle allowed but would imply that
  the neutralino would only be a sub--dominant component of the
  CDM. This constraint will have an extremely strong effect on the
  parameter space of SUSY and allowed ranges of $\mchi$, but not on
  the detection cross sections, as we will see later.


\subsection{Allowed Parameter Space}\label{results:sec}

I will now present the implications of experimental and cosmological
constraints, as described above, on the parameter space of the CMSSM
and the MSO$_{10}$SM.  Next we will discuss the ensuing ranges for
direct detection elastic scattering cross sections.

\begin{figure}[t!]
\begin{center}
\begin{minipage}{12.0cm}
\centerline
{
\hspace*{-.2cm}\psfig{figure=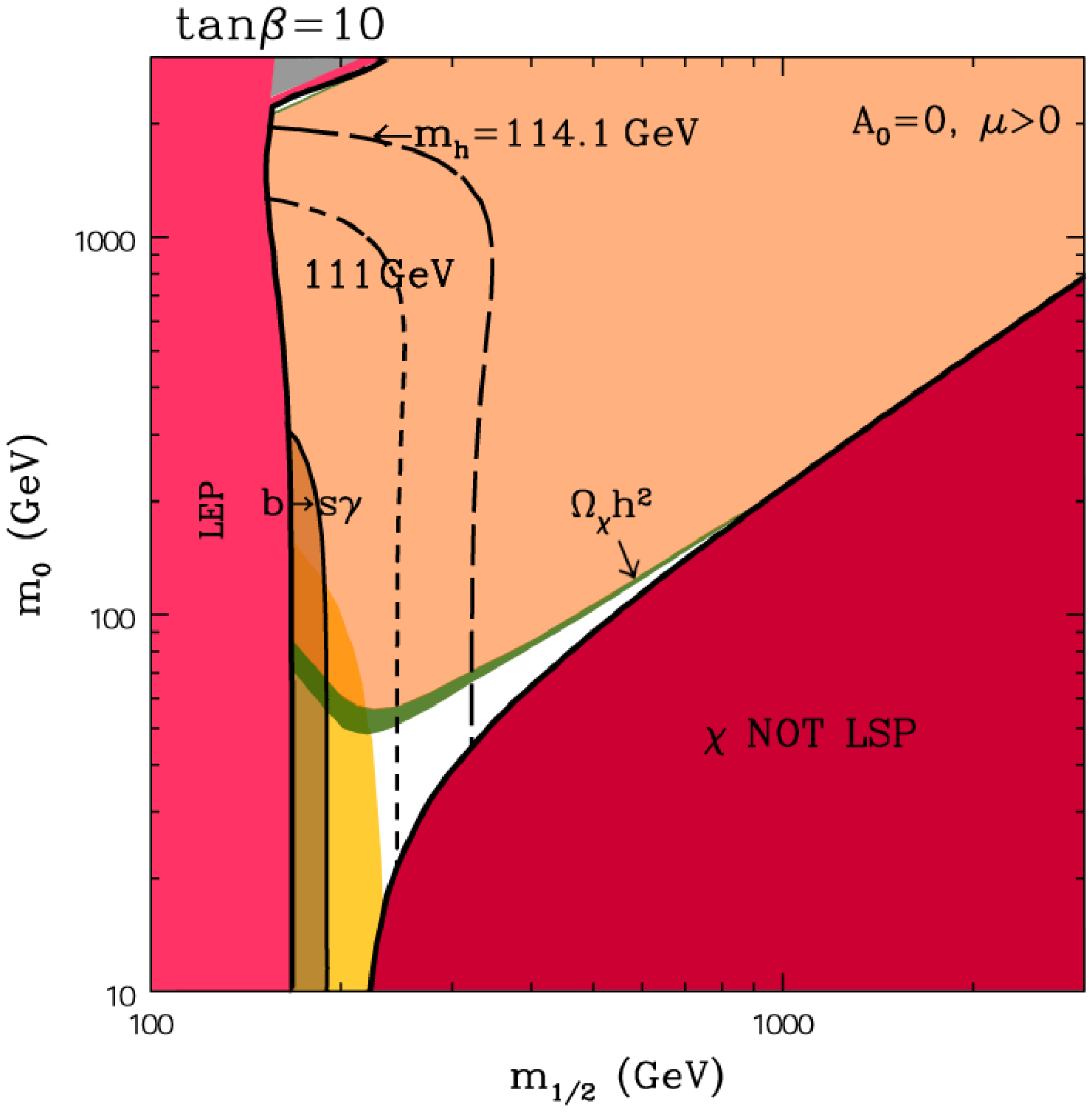, angle=0,width=6.0cm}
\hspace*{-.2cm}\psfig{figure=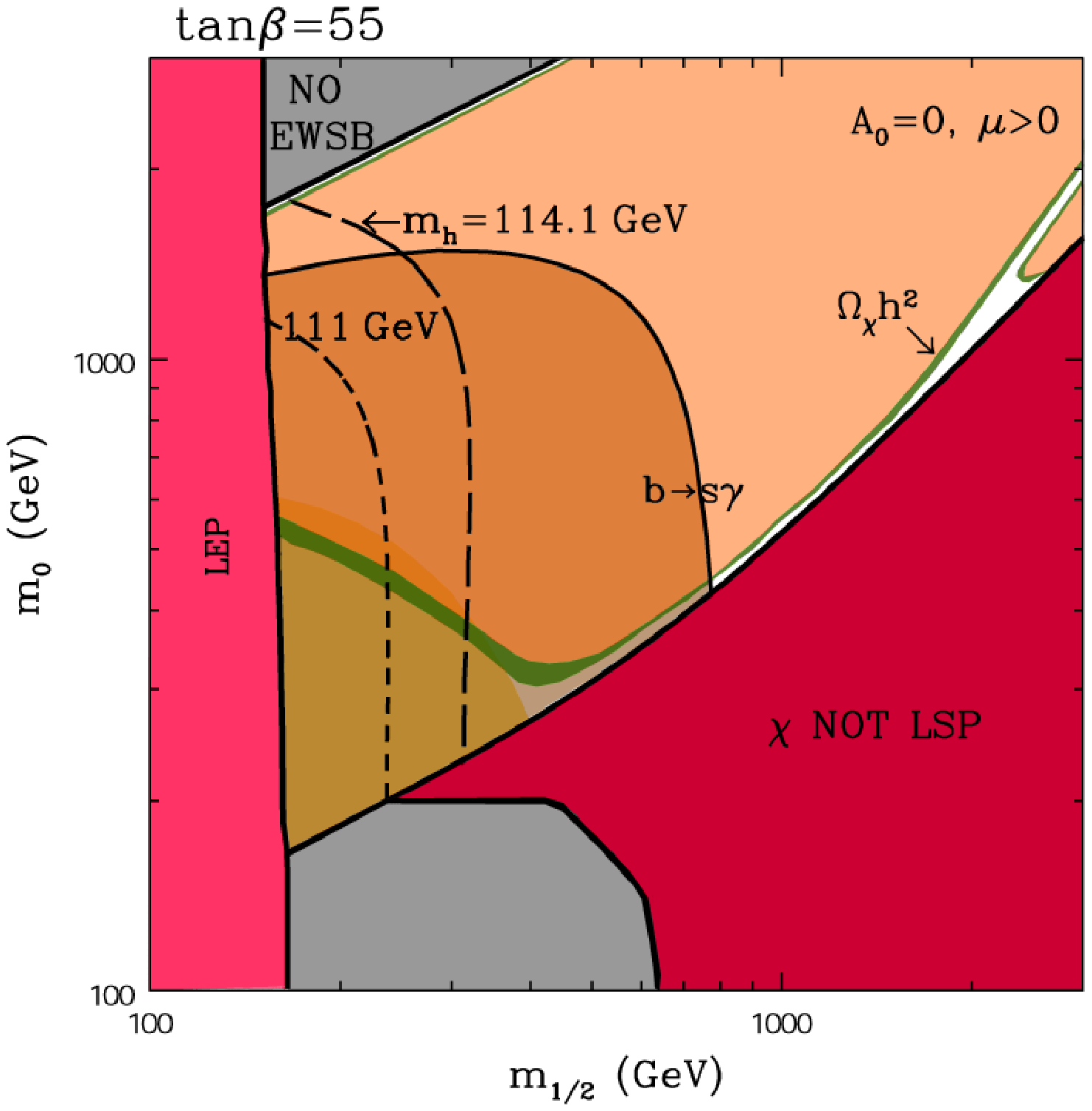, angle=0,width=6.0cm}
}
\end{minipage}
\caption{ The plane ($\protect{\mhalf},\protect{\mzero}$) for
$\azero=0$, $\mu>0$ and for $\protect{\tanb=10}$ (left) and
$\protect{\tanb=55}$ (right). We fix $\mt\equiv\mtpole=175\gev$ and
$\mb\equiv\mbmbsmmsbar=4.25\gev$. The light red bands on the left are
excluded by chargino searches at LEP. In the gray regions the
electroweak symmetry breaking conditions are not satisfied or
$\mha^2<0$ while in the dark red region denoted `$\chi$ NOT LSP' the
LSP is the lighter stau. The large light orange regions of
$\abundchi>0.13$ are excluded by cosmology while the narrow green
bands correspond to the expected range $0.095<\abundchi<0.13$
($2\,\sigma)$. The region to the left of the lightest Higgs scalar
mass $\mhl=111\gev$ and $114.1\gev$ are excluded.  The light brown
region is excluded by $b\rightarrow\gamma$ (assuming MFV).  Also
shown is the semi-oval (yellow) region which is excluded at
$2\,\sigma$ by the anomalous magnetic moment of the muon measurement.
\label{fig:tb1055} }
\end{center}
\end{figure}

\vspace{0.3cm} {\underline{The CMSSM.}}\ \ 
In the CMSSM there are two
distinct cases: low to moderate $\tanb$ regime and the case
$\tanb\sim50$~\cite{rrn1}.  We will illustrate this in
Fig.~\ref{fig:tb1055} in the plane ($\mhalf,\mzero$) for a
representative choice of $\tanb$ for each class. In the left and right
panel of Fig.~\ref{fig:tb1055} the experimental and cosmological
bounds are presented for $\tanb=10$ and $50$, respectively. We also
fix $\azero=0$, $\mu>0$, $\mt\equiv\mtpole=175\gev$ and
$\mb\equiv\mbmbsmmsbar=4.25\gev$. We can see many familiar
features. At $\mhalf\gg\mzero$ there is a dark red wedge where the
$\stauone$ is the LSP. On the other side, at $\mzero\gg\mhalf$ we find
large gray regions where the EWSB is not achieved.  Just below the
region of no--EWSB the parameter $\mu^2$ is small but positive which
allows one to exclude a further (light red) band by imposing the LEP
chargino mass bound. As one moves away from the wedge of no-EWSB,
$\mu^2$ increases rapidly. That implies that, just below the boundary
of the no-EWSB region, the LSP neutralino very quickly becomes the
usual nearly pure bino. This causes the relic abundance $\abundchi$ to
accordingly increase rapidly from very small values typical for light
higgsinos, through the narrow strip ($\Delta\mhalf\sim5\gev$) of the
cosmologically expected (green) range~($0.095<\abundchi<0.13$) up to
larger values which are excluded (light orange). In particular, in the
whole region allowed by the chargino mass bound the LSP is mostly
bino-like.

A general pattern of the cosmologically favored regions does not
change much until $\tanb\sim45-50$ (depending somewhat on $\mb$,
$\mt$, \etc). Generally one finds a robust (green) region of expected
$\abundchi$ at $\mhalf\sim\mzero$ in the range of a few
hundred~$\gev$~\cite{kkrw94:ref}. In addition, at $\mhalf\gg\mzero$,
just above the wedge where the LSP is the $\stauone$, the
coannihilation of the neutralino LSP with $\stauone$ opens up a very
long and  narrow corridor of $\mhalf$ and $\mzero$ favored by
$0.095<\abundchi<0.13$.  At
$\mzero\gg\mhalf$, very close to the region of no-EWSB, again one
finds a very narrow range of $\abundchi$ consistent with observations.


The new feature that appears at large $\tanb$ is the effect of a very
distinct, wide pseudoscalar Higgs resonance in the annihilation
process $\chi\chi\rightarrow A\rightarrow f {\bar f}$. Since, as I
mentioned above, $\mha$ decreases with increasing $\tanb$, at some
point, this opens up a corridor in the plane of
($\protect{\mhalf},\protect{\mzero}$) along $\mha=2\mchi$. Clearly, at
large $\tanb$
cosmological constraints on $\abundchi$  permit much larger
superpartner masses, not only in the very
narrow strips close to the regions of no-EWSB and/or $\stauone$-LSP,
but especially because of the resonance.

The existence of the resonance and of the region of no--EWSB are quite
generic but their exact positions at large $\tanb$ are rather
sensitive to the relative values of the top and bottom
masses. Generally, at fixed $\tanb$, increasing (decreasing) the top
mass relative to the bottom mass causes the region of no--EWSB to move
up (down) considerably because of the diminishing (growing) effect of
the bottom Yukawa coupling on the loop correction to the conditions of
EWSB.  At fixed top and bottom masses, as $\tanb$ decreases, the
region of no--EWSB moves toward somewhat larger $\mzero$ and
smaller $\mhalf$ but the overall effect is not very
significant.

Regarding other constraints, the lightest Higgs mass excludes a
sizable region of smaller $\mhalf$. In the Figures we plot the
contours of $\mhl=114.1\gev$ and of $\mhl=111\gev$ to show the
mentioned earlier effect of the uncertainties in computing $\mhl$.
Larger values of $\mhl$ are given by contours which are shifted along
the $\mhalf$ axis with roughly equal spacings but diverge somewhat at
larger $\mzero$.  The (light brown) region excluded by $\br( B\ra X_s
\gamma )$ grows significantly because the dominant chargino-squark
contribution to the branching ratio grows linearly with $\tanb$.  On
the other hand, one has to remember that the constraint has been
derived in the MFV scenario and can easily be relaxed (or
strengthened) by even a small departure from scheme. Finally,
$(g-2)_{\mu}$ robustly excludes an oval--shape (yellow) region of
small $\mhalf$ and $\mzero$. An upper bound still exists at
$1\,\sigma$ but is now much weaker than before, and it disappears
completely at $2\,\sigma$.

\begin{figure}[t!]
\epsfxsize=8cm
\centerline{\epsfbox{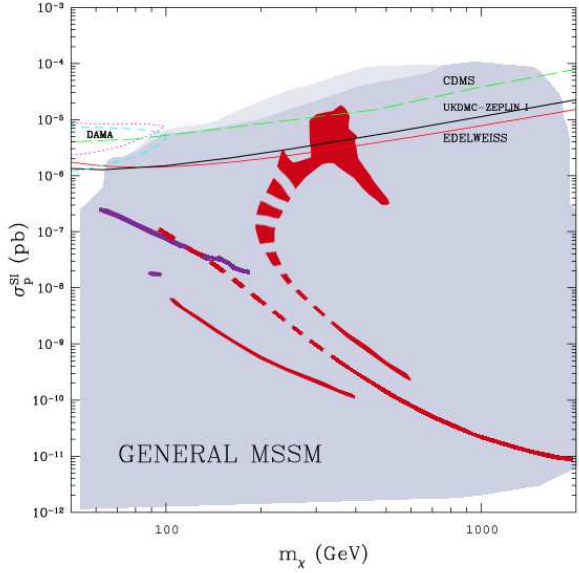}}
\caption{Scalar interaction cross section $\protect\sigsip$ versus
$\protect\mchi$. The lowest (intermediate) red and blue bands
correspond to the cosmologically favored regions of
Fig.~{\protect\ref{fig:tb1055}} with $\tanb=10$ ($\tanb=55$), which
are allowed by experimental constraints, except $(g-2)_\mu$, and by
the favored range $0.095<\abundchi<0.13$ ($2\,\sigma$). The top band
is for the non--unified Higgs model case with $\tanb=60$, $A_0=0$ and
$\delta_u=1$. In the dashed red bands the $\bsgamma$ is not satisfied
(within the MFV scenario). The blue (light blue) region is predicted
by the general MSSM by varying $10<\tanb<65$ and assuming cosmological
and collider constraints including (except for) $\bsgamma$. See text
for more details.  }
\label{sigsipmup:fig}
\end{figure}

One of the most promising strategies to detect WIMPs in the Galactic
halo is to look for the effect of their elastic scattering from a
target material in an underground detector. How do the ``theory plots''
of Fig.~\ref{fig:tb1055} translate into
the more familiar (to the general community) language of WIMP mass
\vs\ cross section ones? And how do they compare with predictions
following from less constrained SUSY models, like the general MSSM?

A relevant quantity is a scalar, or
spin--independent (SI), interaction cross section on a free proton at
zero momentum transfer $\sigsip$. First, in Fig.~\ref{sigsipmup:fig}
we show the case of the CMSSM for the same choices of
parameters as in Figs.~\ref{fig:tb1055}. We plot allowed ranges
of $\sigsip$ \vs\ the neutralino mass. For
comparison, the wide (blue) region is predicted by the general MSSM
assuming rather generous ranges of SUSY
parameters~\cite{knrr1}. In particular, we take $5<\tanb<65$ and
$A_0=0,\pm1\tev$. 

The two lower narrow (red) strips correspond to the allowed
configurations of the CMSSM of Fig.~\ref{fig:tb1055}, with the lower (middle)
one corresponding to $\tanb=10$ ($\tanb=55$). In deriving them, the collider bounds
mentioned above and $0.095\lsim\abundchi\lsim0.13$ were applied, as
well as the $1\sigma$ constraint from $\bsgamma$. Since the last one
is highly sensitive to theoretical assumptions at the unification
scale and excludes large regions of smaller $\mhalf$ and $\mzero$, the
effect of removing it is shown by a dashed line. One can see that, at
large $\tanb$, ranges of much larger $\sigsip$ (at smaller $\mchi$)
become allowed.  On the other hand, we do not apply the constraint
from $\amu$. Its effect in the right window would be to exclude (at
$2\,\sigma$) the part with $\mchi\lsim150\gev$ in the main (red) band,
but {\em not} in the (violet) focus point band. For $\tanb=10$ instead
the Higgs mass bound remains stronger.

For each $\tanb$ a general pattern is that of two distinct
``branches''. The lower (red) one corresponds for the most part to the
$\mha$--resonance and/or ${\widetilde \tau}$--coannihilation region at
$\mhalf\gg\mzero$ in Figs.~\ref{fig:tb1055}. On the other hand, the
upper (violet) branch comes from the focus point region of large
$\mzero\gg\mhalf$. In this region, even though the masses of squarks
and sleptons are large, in the~\tev\ range, the mass of the lightest
neutralino remains moderate, since roughly $\mchi\sim0.4\mhalf$. In
both branches, the dominant contribution to $\sigsip$ comes from the
exchange of the heavy scalar Higgs.

One can clearly see that predictions of the CMSSM are rather
definite (thus fully justifying the name the model bears). In
particular, note a remarkably narrow bands of $\sigsip$ as a function
of $\mchi$. Generally larger values of $\sigsip$ and smaller $\mchi$
are predicted by the {\em extremely narrow} focus point region of
$\mzero\gg\mhalf$, while the long (red) tail comes from the region of the
neutralino--slepton coannihilation and (at large $\tanb\gsim50$) of
the wide resonance. Generally, as $\tanb$ increases, so does the cross
section at fixed $\mchi$.  

This implies that, for both small and large $\tanb$ there is also an
{\em upper} bound on $\mchi$ because the cosmologically allowed
regions in the plane ($\mhalf,\mzero$) eventually end at large
$\mhalf\gg\mzero$. As $\mhalf$ grows, the neutralino--slepton coannihilation eventually
ceases to remain effective and $\abundchi$ grows beyond $0.13$. At very
large $\tanb\gsim50$ the $A$--resonance extends in the plane
($\mhalf,\mzero$) up to very large values of $\mhalf$ but eventually
ends, too. Thus follows a {\em lower} bound on $\sigsip$ of about
$\sim8\times10^{-11}\pb$ and an {\em upper} limit
$\mchi\lsim1950\gev$, as can be (barely) seen in the Figure.

The top red region (with an amusing shape) has been added to show a
rather exceptional case of boundary conditions for which the predicted
$\sigsip$ is unusually high, in fact already partly excluded by
experiment.  It corresponds to the scenario where the soft mass terms
of the Higgs doublet at the GUT scale is not unified with the other
scalars. The case presented in the Figure is for $\tanb=60$, $A_0=0$
and $\delta_u=1$ where
$m^2_{H_2}=\mzero^2\left(1+\delta_u\right)$. This specific example is
somewhat unusual because for other choices of $\tanb$ and/or
$\delta_u<1$ the predicted ranges of $\sigsip$ are typically
$\lsim10^{-7}\pb$, like in the CMSSM, while on the other hand usually
giving patterns distinctively different from those predicted in the
CMSSM. The hashed (full) red region corresponds to relaxing (imposing)
the constraint from $b\ra s\gamma$.

In contrast to the specific patterns resulting from different
unification assumptions, in the general MSSM one is struck by the
enormity of the {\em a priori} allowed ranges, extending down to $\sim
10^{-12}\pb$, or, at lower $\mchi$, even below~\cite{knrr1}.  The lowest values of
$\sigsip$ generally correspond to one or more SUSY mass parameters
being very large, above $1\tev$, and thus can be considered more
fine--tuned, and therefore (hopefully!) less likely. Requiring SUSY
mass parameters to be less than a few~\tev\ allows one to put a lower
bound on $\sigsip$.

At lower $\mchi$ the lower bound on $\sigsip$ can actually extend to
much lower values. This is because $\abundchi$ is determined there
primarily by neutralino coannihilation with sleptons. By selecting a
slepton mass not too much above $\mchi$, one can make
coannihilation reduce $\abundchi$ to the favored range and have at the
same time very low $\sigsip$. This requires some fine--tunning but
three or four orders of magnitude for $\sigsip$ (or even smaller)
become allowed by allowing for SUSY mass parameters in the multi--TeV
range. The coannihilation with sleptons is very effective at $\mchi$
in the range of several hundred GeV (with the upper limit growing with $\tanb$) but at
some point ceases to be effective enough and $\abundchi$ grows to too
large values. One can still reduce it to acceptable values by either
fine--tuning $\mha$ to twice $\mchi$ or by taking $|\mu|$ very (!)
large.

That's for the lower ranges, which hopefully will never have to be
probed! On the upper side, values as large as $\sim 10^{-5}\pb$ are
allowed. At lower $\mchi$ they are limited from above by a lower bound
on light Higgs mass and by the lower limit on $\abundchi$. At larger $\mchi$ the
limit from $\bsgamma$ is the main constraint. The effect of lifting it
is marked by a lighter shade of blue. The constraint from
$\amu$ puts an upper bound (at $1\,\sigma$) of $\mchi\gsim450\gev$ but
not on $\sigsip$~\cite{knrr1}, and not  at $2\,\sigma$.

The current experimental sensitivity is also shown for
comparison. Denoted are the regions claimed by DAMA to be consistent
with the annual modulation signal allegedly present in their data when
one ignores (magenta dots) or includes (blue dash) the Collaboration's
previous limits. Some other experiments (CDMS, UKDMC, Edelweiss) have
by now excluded much of the DAMA region. 

\begin{figure}[t!]
\begin{center}
\begin{minipage}{12.0cm}
\centerline
{\hspace*{-.2cm}\psfig{figure=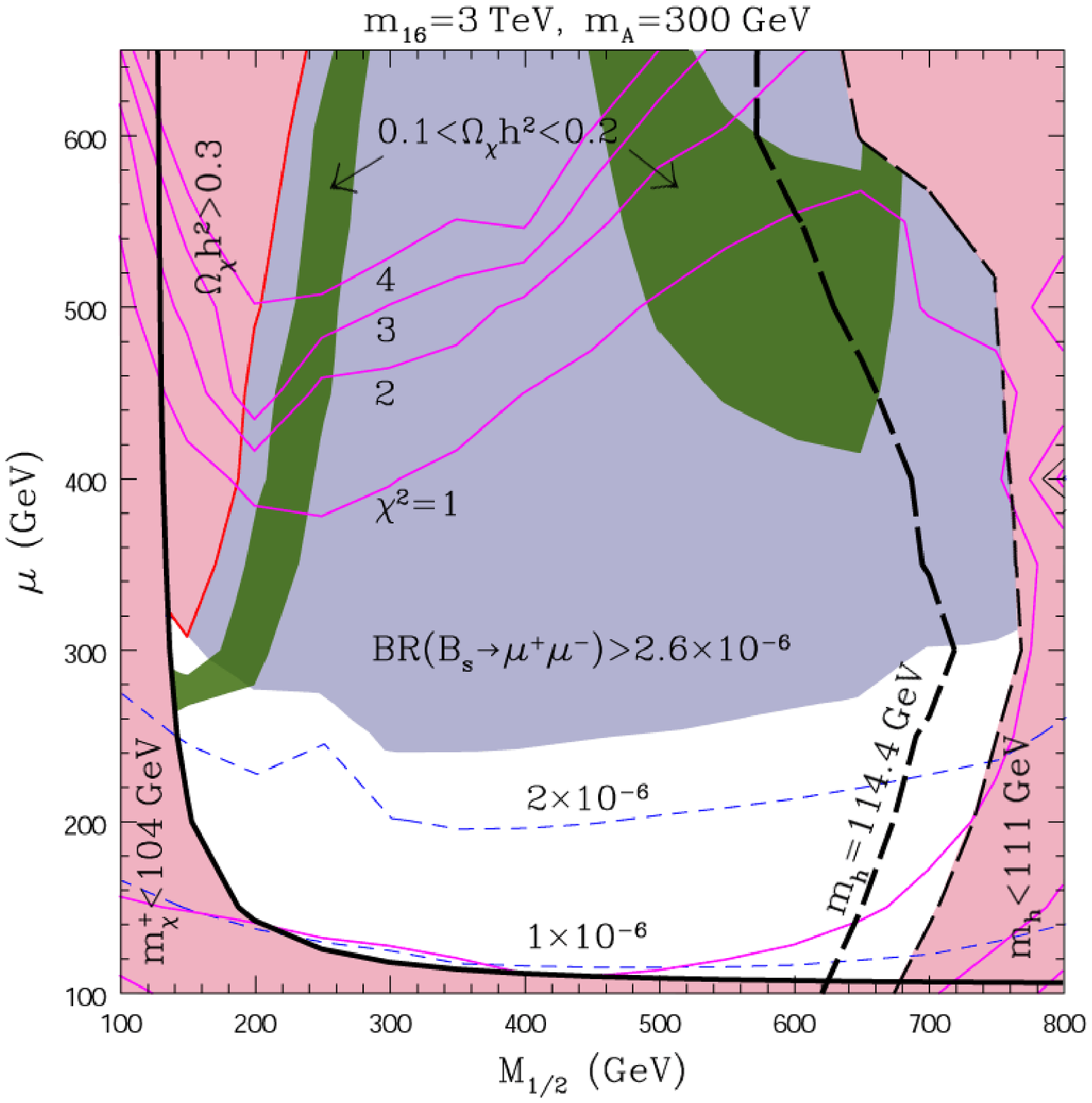, angle=0,width=6.0cm}
\hspace*{-.2cm}\psfig{figure=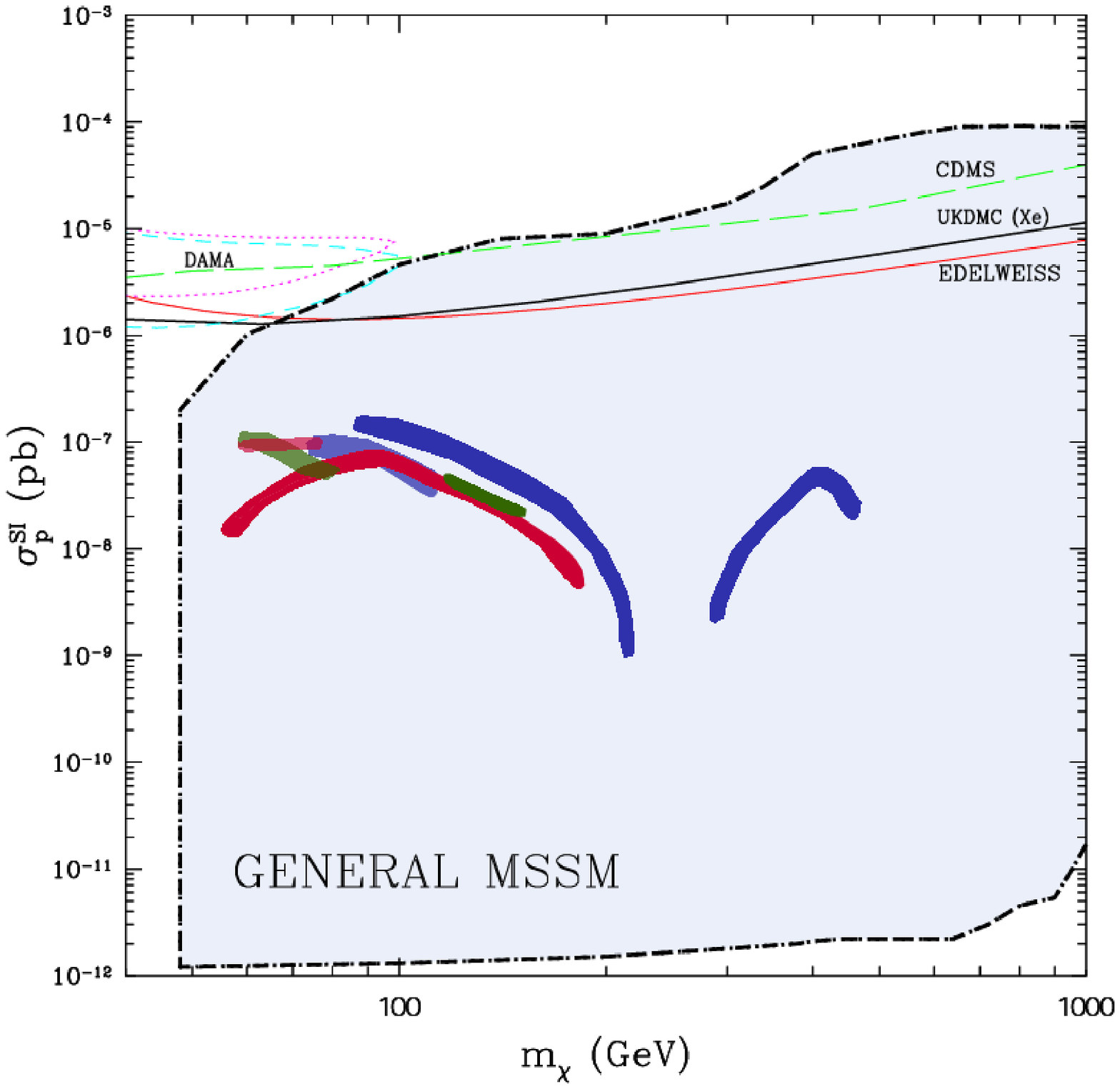, angle=0,width=6.0cm}
}
\end{minipage}
\caption{Left panel:~The cosmologically favored (green) and allowed (white) regions
  in MSO$_{10}$SM for $m_{16}=3\tev$ and $\mha=300\gev$. Red and blue regions
  are excluded, as marked in the panel. Right panel:~Allowed regions
  of $\sigsip$ in MSO$_{10}$SM for different choices of
  parameters. See text for more details.
\label{fig:so10}}
\end{center}
\end{figure}

\vspace{0.3cm}
\noindent{\underline{The MSO$_{10}$SM.}}\ \
A typical example of the parameter space is presented in
the left panel of Fig.~\ref{fig:so10}. (For other cases
see~\cite{drrr1} and the talk by Raby~\cite{raby_pascostalk}.) The
mass of the pseudoscalar is fairly low which allows for efficient
depletion of the number of WIMP neutralinos in the early Universe
through the $A$--resonance, as in the case of the CMSSM.  The vertical
position
of the pole is clearly visible in the panel at $\mhalf\simeq350\gev$. Away from it,
$\abundchi$ is too large, close to it it's too small. In-between one
finds sizable cosmologically favored (green) regions. Impact of
other relevant constraints is also shown. We do not apply the
constraint from $b\to s\gamma$ because because in this model one is
not bound by the MFV framework. On the other hand, all the sfermions
are heavy and the model predicts basically the SM value for
$(g_\mu-2)/2$.

Another aspect of the MSO$_{10}$SM which is of much interest to
phenomenology is the process $B_s\rightarrow \mu^+ \mu^-$. Because the
process involves a flavor--changing exchange of the pseudoscalar,
which is fairly light, the resulting $\brbsmumu$ can be large, often
well within the reach of Run~II at the Tevatron. At lower $\mha$, it
can even exceed the current bound $2.6 \times 10^{-6}$ from CDF which
is expected to be improved by a some two orders of magnitude. In
short, the process offers very good prospects for the Tevatron. 


In the right panel of Fig.~\ref{fig:so10} we present $\sigsip$ \vs\
$\mchi$ for several typical choices of $\mha=300\gev$ (light shade)
and $500\gev$ (dark shade), and for $m_{16}=2.5\tev$ (green), $3\tev$ (red)
and $5\tev$ (blue)~\cite{drrr1}. The patters are somewhat different
from those of the CMSSM (thus possibly allowing for discriminating
between the two models should a WIMP signal be detected and $\sigsip$
and $\mchi$  measured with some accuracy). One also finds
$\sigsip\lsim10^{-7}\pb$ although also $\sigsip\gsim10^{-9}\pb$ which
is encouraging, compared to the CMSSM.

In summary, it is clear that the current experimental sensitivity is
sufficient to already probe a part of the parameter space of the
general MSSM. On the other hand, it is still typically at least one
order of magnitude above the preferred ranges of cross sections that
are predicted by unified models as the examples of the CMSSM and the
MSO$_{10}$SM demonstrate.  For these ranges to be explored a new
generation of detectors will be required and is actually already being
constructed.  On the other hand, theoretically lower ranges of WIMP
mass and therefore larger $\sigsip$ are more natural (fine tuning and
$\amu$). It is therefore quite possible that a nice surprise may come
before long. Finally, since the specific ranges of $\sigsip$ and
$\mchi$ are typically very narrow and model dependent, once a WIMP
signal is detected, one may hope to be able to discriminate among
different unification scenarios.

\vspace*{0.5cm}
I am indebted to Professor D.P.~Roy for his kind invitation and to the
whole Organizing Committee for preparing a very successful and
interesting meeting.


\begin{thebibliography}{99}

\bibitem{wmap0302} D.~N.~Spergel {\it
et al.},
Astrophys. J. Suppl. {\bf 148} (2003) 175. 

\bibitem{leeweinberg77} 
B.W.~Lee and S.~Weinberg, Phys. Rev. Lett. {\bf 39} (1977) 165-168. 
For more references, see
E.W.~Kolb and M.S.~Turner,
The Early Universe, Addison-Wesley, Redwood City, 1990.

\bibitem{gk90}
K.~Griest and M.~Kamionkowski, Phys. Rev. Lett. {\bf 64} (1990) 615.

\bibitem{rtw90}
K.~Rajagopal, M.S.~Turner and F.~Wilczek, Nucl. Phys. {\bf B 358} (1991) 447.

\bibitem{ckrplus}
L.~Covi, J.E.~Kim and L.~Roszkowski, Phys. Rev. Lett. {\bf 82} (1999) 4180;
L.~Covi, H.B.~Kim, J.E.~Kim and L.~Roszkowski, \jhep{0105}{2001}{033};
L.~Covi, L.~Roszkowski and M.~Small, \jhep{0207}{2002}{023}.

\bibitem{gravitinoproduction}
H. Pagels and J.R. Primack, 
\prl{48}{1982}{223};
S. Weinberg, Phys. Rev. Lett. \prl{48}{1982}{1303}; 
J. Ellis, A.D.~Linde and D.V. Nanopoulos, \plb{118}{1982}{59}; 
\plb{443}{1998}{209}; and several more recent papers.

\bibitem{wimpzilla:ref} 
D.J.H.~Chung, E.W.~Kolb and A.~Riotto, \prl{81}{1998}{4048};
V.~Kuzmin and I.~Tkachev, hep-ph/9809547.

\bibitem{st02}
G.~Servant and T.~Tait, 
\npb{650}{2003}{391}  
and 
New J. Phys. {\bf 4} (2002) 99. 


\bibitem{mohapatra:mirrordm} 
R.N.~Mohapatra and V.L.~Teplitz,
\prd{62}{2000}{063506}. 

\bibitem{raby_pascostalk} S.~Raby, talk at this  conference.

\bibitem{nojiri_pascostalk} M.~Nojiri, talk at this  conference.

\bibitem{bdr}  T. Bla\v{z}ek, R. Derm\' \i \v sek and S. Raby,
\prl{88}{2002}{111804}; \prd{65}{2002}{115004}.

\bibitem{drrr1} R. Derm\' \i \v sek, S. Raby, L. Roszkowski and R. Ruiz de
 Austri, \jhep{0304}{2003}{037}.

\bibitem{dn93}
M. Drees and M. Nojiri, \prd{47}{1993}{376}.

\bibitem{jkg96:ref}
See, e.g., G. Jungman, M. Kamionkowski and K. Griest, 
\prep{267}{1996}{195}.

\bibitem{chiasdm} L.~Roszkowski, Phys. Lett. {\bf B 262} (1991) 59;
see also J.~Ellis, D.V.~Nanopoulos, L.~Roszkowski, and D.N.~Schramm, 
Phys. Lett. {\bf B 245} (1990) 251.

\bibitem{na92} P.~Nath and R.~Arnowitt, Phys. Lett. 
{\bf B 289} (1992) 368.

\bibitem{rr93} R.G.~Roberts and L.~Roszkowski, Phys. Lett. 
{\bf B 309} (1993) 329.

\bibitem{kkrw94:ref} G.L.~Kane, C.~Kolda, L.~Roszkowski, and J.~Wells, 
Phys. Rev. {\bf D 49} (1994)  6173.

\bibitem{rrn1}
L.~Roszkowski, R.~Ruiz de Austri and T.~Nihei,
\jhep{0108}{2001}{024}. 

\bibitem{or1+2} K.~Okumura and L.~Roszkowski, 
hep-ph/0208101 and \jhep{0310}{2003}{024}.

\bibitem{bcmu02}
A.~J.~Buras, A.~Czarnecki, M.~Misiak and J.~Urban,
\npb{631}{2002}{219}.


\bibitem{dehz03} 
M.~Davier, S.~Eidelman, A.~Hocker and Z.~Zhang, hep-ph/0308214.

\bibitem{knrr1} Y.~G.~Kim, T.~Nihei, L.~Roszkowski and R.~Ruiz de Austri,
\jhep{0212}{2002}{034}. 

\end{thebibliography}
\end{document}